\documentclass[11pt,a4paper]{article}
\usepackage{jheppub}

\newcommand{\ket}[1]{|#1\rangle}
\newcommand{\rket}[1]{|#1)}
\newcommand{\rkets}[1]{\left|#1\right)}
\newcommand{\inp}[2]{\langle#1|#2\rangle}
\newcommand{\rinp}[2]{(#1|#2)}
\newcommand{\matens}[3]{\langle#1|#2|#3\rangle}
\newcommand{\rmate}[3]{(#1|#2|#3)}
\newcommand{\expv}[1]{\langle#1\rangle}
\newcommand{\comm}[2]{[#1,#2]}
\newcommand{\acomm}[2]{\{#1,#2\}}
\newcommand{\matcc}[2]{\left[\begin{array}{cc} #1 \\ #2 \end{array}\right]}
\newcommand{\matccc}[3]{\left[\begin{array}{ccc} #1 \\ #2 \\ #3 \end{array}\right]}
\newcommand{\eref}[1]{\eqref{#1}}
\newcommand{\kz}{\hat{K}_0}
\newcommand{\kp}{\hat{K}_+}
\newcommand{\km}{\hat{K}_-}
\newcommand{\xop}{\hat{x}}
\newcommand{\pop}{\hat{p}}
\newcommand{\hop}{\hat{H}}
\newcommand{\dop}{\hat{D}}
\newcommand{\kop}{\hat{K}}
\newcommand{\vop}{\hat{V}}
\newcommand{\ttop}{\hat{T}}
\newcommand{\ddt}[1]{\frac{d#1}{dt}}
\newcommand{\ddtt}[1]{\frac{d^2#1}{dt^2}}
\newcommand{\prm}{\hspace{-0.09cm}'}
\newcommand{\oforder}[1]{\mathcal{O}(#1)}
\newcommand{\hs}[1]{\hspace{#1 cm}}
\newcommand{\hcal}{\mathcal{H}}

\newcommand{\scal}{\mathcal{S}}
\newcommand{\partialsi}[1]{\partial_{#1}}

\title{Duality constructions from quantum state manifolds}
\author[a]{J.N. Kriel,}
\author[a]{H.J.R. van Zyl}
\author[a,b]{and F.G. Scholtz}
\affiliation[a]{Institute of Theoretical Physics, Stellenbosch University, Stellenbosch 7600, South Africa}
\affiliation[b]{National Institute for Theoretical Physics (NITheP), Stellenbosch 7600, South Africa}
\emailAdd{hkriel@sun.ac.za}
\emailAdd{hvanzyl@sun.ac.za	}
\emailAdd{fgs@sun.ac.za}
\keywords{Holography and condensed matter physics (AdS/CMT), AdS-CFT Correspondence, 2D Gravity}

\abstract{The formalism of quantum state space geometry on manifolds of generalised coherent states is proposed as a natural setting for the construction of geometric dual descriptions of non-relativistic quantum systems. These state manifolds are equipped with natural Riemannian and symplectic structures derived from the Hilbert space inner product. This approach allows for the systematic construction of geometries which reflect the dynamical symmetries of the quantum system under consideration. We analyse here in detail the two dimensional case and demonstrate how existing results in the $AdS_2/CFT_1$ context can be understood within this framework. We show how the radial/bulk coordinate emerges as an energy scale associated with a regularisation procedure and find that, under quite general conditions, these state manifolds are asymptotically anti-de Sitter solutions of a class of classical dilaton gravity models. For the model of conformal quantum mechanics proposed by de Alfaro et. al. \cite{alfaro_conformal_1976} the corresponding state manifold is seen to be exactly $AdS_2$ with a scalar curvature determined by the representation of the symmetry algebra. It is also shown that the dilaton field itself is given by the quantum mechanical expectation values of the dynamical symmetry generators and as a result exhibits dynamics equivalent to that of a conformal mechanical system.}

\begin{document}
\maketitle
\section{Introduction}
\label{sectionintroduction}
Geometric structures have come to play an increasingly prominent role in the development of quantum theory. An early instance of this is Kibble's \cite{kibble_geometrization_1979} notion of the \emph{geometrisation of quantum mechanics}, based on the observation \cite{chernoff1974some} that the (projective) Hilbert space is equipped with a natural symplectic structure in terms of which quantum dynamics can be recast in a form identical to that of Hamilton's formulation of classical mechanics. Following this, Provost and Vallee \cite{provost_riemannian_1980} highlighted that quantum state space also permits a Riemannian metric, compatible with the symplectic structure, and analysed its properties on submanifolds of the Hilbert space consisting of generalised coherent states \cite{perelomov_generalized_1986}. A range of studies have subsequently contributed to the formalism and applications of quantum state space geometry. See e.g. \cite{brody_geometric_2001} and references therein. Of interest here is the work of Ashtekar and Schilling \cite{ashtekar_geometrical_1997}, which provides a detailed analysis of the physical content these structures from the unifying viewpoint of K\"ahler geometry. Here we will investigate applications of state space geometry to another topic of much current interest, which shares a strong geometric underpinning. The $AdS/CFT$ correspondence \cite{maldacena_large_1999,witten_anti_1998, witten_anti-sitter_1998, gubser_gauge_1998} is a conjectured duality between $d+1$-dimensional quantum gravity on anti-de Sitter space ($AdS$) and a conformal field theory ($CFT$) defined on the $d$-dimensional boundary of $AdS$. Central to this duality is the matching of the symmetries of the two theories: the conformal group in $d>2$ spacetime dimensions is exactly  the isometry group of $d+1$ dimensional $AdS$. A large body of literature has developed around applying the principles of $AdS/CFT$ to the condensed matter context \cite{hartnoll_lectures_2009,mcgreevy_holographic_2010,sachdev_condensed_2011}, and here too the matching of the symmetries of the quantum system with those of the geometry is a key feature \cite{balasubramanian_gravity_2008-1,son_toward_2008}.\\

In this paper we highlight the usefulness of state space geometry as a means of systematically constructing  geometric dual descriptions of quantum systems. Several characteristic features of the $AdS/CFT$ picture will be seen to emerge quite naturally in this approach. The central ideas are as follows:
\begin{enumerate}
	\item The Hilbert space $\mathcal{H}$ of a quantum mechanical system is equipped with a natural metric $g$ and symplectic form $\sigma$. Within $\mathcal{H}$ we construct a set $\mathcal{S}$ of generalised coherent states,  generated through the action of the system's dynamical symmetry group on a fixed reference state. In contrast to most standard coherent state constructions it will be useful to consider a reference state which is of infinite norm. The elements of $\mathcal{S}$ are therefore more accurately described as members of the rigged Hilbert space containing $\mathcal{H}$. By construction the action of the dynamical symmetry group leaves $\mathcal{S}$ invariant.
	\item The system's Hamiltonian is an element of the dynamical symmetry algebra. As a result, one of the parameters labelling the elements of $\mathcal{S}$ has the natural interpretation of a time coordinate. The inner product of two such states can then be regarded as a transition amplitude which will exhibit transformation properties which reflect the  system's dynamical symmetries. In fact, this transition amplitude will act as the potential function, similar to a K\"ahler potential, from which the geometry is generated.
	\item Since the elements of $\mathcal{S}$ are not normalisable they do not belong to the Hilbert space proper, and the metric $g$ and symplectic form $\sigma$ are therefore not defined on $\mathcal{S}$ itself. To remedy this we must regularise the elements of $\mathcal{S}$ to render them normalisable while preserving the invariance of the resulting set of states under the dynamical symmetry group. This is done by introducing additional regularisation coordinates, thereby also increasing the dimensionality of the state manifold. These additional dimensions are analogues of the radial coordinate appearing in the standard $AdS/CFT$ picture and has, for the examples considered here, the interpretation of an energy scale. The resulting state manifold is partitioned into bulk and boundary regions, consisting of finite and infinite norm states respectively. The metric and symplectic forms is now defined within the bulk with the dynamical symmetries acting as isomorphisms for both these structures.
	\item Drawing on ideas from the geometric formulation of quantum mechanics it is shown how quantum mechanical expectation values are related to scalar and vector fields on the state manifold. In particular, the dynamical symmetry generators translate into real scalar fields of which the Hamiltonian vector fields are precisely the Killing fields. Conformal transformations are also considered in this framework.
\end{enumerate}
Although this program is quite general we will focus here on the two dimensional case, having in mind applications to the $AdS_2/CFT_1$ correspondence. Various perspectives on this duality have appeared in the literature. On the $CFT_1$ side  the model of conformal quantum mechanics ($CQM$) proposed by de Alfaro et. al. \cite{alfaro_conformal_1976} has been studied in this context \cite{chamon_conformal_2011,cadoni_2d_2001,jackiw_conformal_2012-1,molina-vilaplana_xp_2013} while on the gravity side models of dilaton gravity have featured prominently \cite{cadoni_asymptotic_1999,cadoni_symmetry_2000,strominger_ads2_1999}. See also \cite{cadoni_ads/cft_2001} and references therein. Not surprisingly, the symmetries of $AdS$ spacetime are again found to be closely related to those of $CQM$. The role that coherent states and the symplectic structure on $AdS_2$ plays within this duality have recently also been studied by Axenides et. al. \cite{axenides_modular_2014}.

In this two-dimensional setting the framework above also allows for a gravitational perspective on the state manifold:
\begin{enumerate}
	\item[5.] Under quite general conditions the two dimensional state manifolds that emerge from this construction are found to be asymptotically $AdS$ solutions of a class of classical dilaton models. In the case of $CQM$ the metric is found to be exactly $AdS$ and the infinite norm boundary states coincide with those identified in Chamon et. al. \cite{chamon_conformal_2011}. The dilaton field is shown to correspond to the expectation values of the dynamical generators. Furthermore, the Heisenberg picture dynamics of these generators are imprinted on the geometry and thereby also on the dilaton equations of motion. As a result the dilaton dynamics are very closely linked to that of  a conformally invariant quantum system, a result previously observed in \cite{cadoni_2d_2001}. 
\end{enumerate}
  
The paper is organised as follows. In section \ref{sectiongeo} we collect results pertaining to geometric structures on quantum state manifolds and show how the symmetries of these structures relate to those of the system under consideration. Here we also establish the link between quantum mechanical expectation values and Hamiltonian flows on the manifold. Section \ref{section1d2d} is dedicated to an analysis of two dimensional geometries which result from complexified time evolution. In section \ref{sectionCQM} we focus on the de Alfaro-Fubini-Furlan model of $CQM$ and show how an $AdS_2$ geometry emerges here. We also consider the generators of conformal transformations, both on the operator and geometric levels, and derive their equations of motion.  In section \ref{sectiondilaton} we adopt a gravitational perspective and show how the state manifold emerges as the solution of classical dilaton gravity. Here the link between the symmetry generators and the dilaton itself is established. In section \ref{sectionconclusion} we conclude and identify some avenues for further investigation. Some technical results are collected in the appendix.

\section{Geometric structures on quantum state manifolds}
\label{sectiongeo}
\subsection{Definitions}
\label{sectionProvost}
We begin by outlining the construction of Provost and Vallee \cite{provost_riemannian_1980} for defining a natural Riemannian structure on a manifold of quantum states contained in a Hilbert space $\hcal$. Consider a family of normalised state vectors $\scal=\{\ket{s}\}\subseteq\hcal$ parametrized smoothly by a set of coordinates $s=(s_1,s_2,\ldots,s_n)\in\mathbb{R}^n$. Associated with each $\ket{s}$ is a ray $\widetilde{\ket{s}}$ representing the set $\{e^{i\theta}\ket{s}\,:\,\theta\in\mathbb{R}\}$ of physically equivalent state vectors. The goal is to define two geometric structures on the state manifold $\scal$, a metric tensor $g$ and a closed 2-form $\sigma$. Both of these will be derived from the inner product already defined on $\hcal$. For these structures to have physical significance requires that they are invariant under the replacement $\ket{s}\rightarrow e^{i\phi(s)}\ket{s}$ which modifies the vectors but not the physical content of the states they represent. If this condition is met both $g$ and $\sigma$ may be regarded as being defined on the manifold of rays $\widetilde{\scal}=\{\widetilde{\ket{s}}\}$.\\

Following \cite{provost_riemannian_1980} we define
\begin{equation}
	\beta_j(s)\equiv-i\partial_j\prm \inp{s}{s'}|_{s=s'}\quad{\rm and}\quad\gamma_{ij}(s)+i\sigma_{ij}(s)\equiv\partial_i\partial_j\prm\inp{s}{s'}|_{s=s'}
	\label{provost1}
\end{equation}
where $\partial_i\equiv\frac{\partial}{\partial s_i}$. The tensors $\gamma_{ij}$ and $\sigma_{ij}$ correspond to the real and imaginary parts of the inner product and satisfy $\gamma_{ij}=\gamma_{ji}$ and $\sigma_{ij}=-\sigma_{ji}$. Now  $\sigma=\sigma_{ij}ds_i\wedge ds_j$ while the metric tensor is defined as
 \begin{equation}
	g_{ij}(s)=\gamma_{ij}(s)-\beta_i(s)\beta_j(s).
	\label{provost2}
\end{equation}
The subtraction of the $\beta_i(s)\beta_j(s)$ terms ensures the invariance of $g_{ij}(s)$ under the replacement \mbox{$\ket{s}\rightarrow e^{i\phi(s)}\ket{s}$}. This invariance, for both $g$ and $\sigma$, is apparent in the equivalent definitions
\begin{equation}
	g_{ij}(s)=[\,\partial_i\partial_j\prm\log|\inp{s}{s'}|\,]_{s=s'}\quad{\rm and}\quad \sigma_{ij}(s)=\frac{1}{2i}\left[\,\partial_i\partial_j\prm\log\frac{\inp{s}{s'}}{\inp{s'}{s}}\,\right]_{s=s'}.
	\label{provost3}
\end{equation}
These expressions also have the benefit of being invariant under general rescalings $\ket{s}\rightarrow N(s)\ket{s}$ where the scalar function $N(s)$ need not be a phase. In what follows we use the notation $\rket{\cdot}$ to denote a potentially unnormalised state. The definitions in \eref{provost3} may therefore be applied directly to a family of such states $\{\rket{s}\}$ by replacing $\inp{s}{s'}$ by $\rinp{s}{s'}$. This approach is often  computationally convenient.\\

Of particular interest are manifolds of unnormalised states $\scal=\{\rket{z}\}$ which are parametrized holomorphically by a set of complex coordinates $z=(z_1,\ldots,z_m)\in\mathbb{C}^m$. The inner product $\rinp{z}{z}$ then depends on $z$ and $\bar{z}$ through the ket and bra respectively. In these coordinates the non-zero components of $g$ and $\sigma$ read
\begin{equation}
	g_{a\bar{b}}=g_{\bar{b}a}=\frac{1}{2}\partial_a \partial_{\bar{b}}\log \rinp{z}{z}\hs{1}{\rm and}\hs{1}
	\sigma_{a\bar{b}}=-\sigma_{\bar{b}a}=ig_{a\bar{b}}
	\label{generalkahler}
\end{equation}
where $\partial_a\equiv\frac{\partial}{\partial z_a}$, $\partial_{\bar{a}}\equiv\frac{\partial}{\partial \bar{z}_a}$ and $a,b\in\{1,\ldots,m\}$. Also, the 2-form $\sigma$ will be nondegenerate and therefore a symplectic form, and so $\scal$ is a K\"ahler manifold \cite{moroianu_lectures_2004} with K\"ahler potential $\frac{1}{2}\log\rinp{z}{z}$. 

\subsection{Dynamical symmetries}
\label{generalisometries}
Since $g$ and $\sigma$ are derived from the inner product on $\mathcal{H}$ these structures should be invariant under unitary transformations that leave the manifold of rays $\widetilde{\scal}$ invariant. If $\hat{U}$ is such a transformation it will induce on $ \scal$ a mapping $s\rightarrow u(s)$ according to 
\begin{equation}
	\hat{U}\ket{s}=e^{i\phi(s)}\ket{u(s)}
	\label{generalaction}
\end{equation}
with $\phi(s)$ a scalar function. In the applications that follow $\hat{U}$ will represent a dynamical symmetry of the system under consideration. We therefore adopt the same terminology here, and refer to these transformations as \emph{dynamical symmetries of $\scal$.} As a result of \eref{generalaction} the inner product of two elements of $\scal$ will obey
\begin{equation}
	\inp{s}{s'}=e^{-i\phi(s)}e^{i\phi(s')}\inp{u(s)}{u(s')}.
	\label{generalinp}
\end{equation}
Abbreviating $t=u(s)$ and using the definition of the metric in \eref{provost3} now reveals that
\begin{equation}
	g_{ij}(s)=\frac{\partial t_k}{\partial s_i}\frac{\partial t_l}{\partial s_j}g_{kl}(t)
\end{equation}
or, compactly, that $g=u^*g$. The mapping $s\rightarrow u(s)$ is therefore an isometry of the metric and, by a similar argument, also leaves $\sigma$ invariant. This link between the dynamical symmetries on the quantum mechanical and geometric levels is discussed further in section \ref{sectionremarks}.

\subsection{Manifolds generated from group actions}
\label{sectionGCS}
We will be interested mainly in manifolds of generalised coherent states which are generated through the action of a Lie group on a fixed reference state \cite{perelomov_generalized_1986,klauder1985applications,zhang_coherent_1990,twareque_ali_coherent_1995}. Here we outline the basic construction, following closely the approach of \cite{perelomov_generalized_1986}. Let $G$ be a Lie group with a unitary representation $T(g)$ acting on a Hilbert space $\hcal$ and let $\ket{\psi_0}\in\hcal$ be an arbitrary fixed reference state. The action of $G$ on $\ket{\psi_0}$ then generates the set of states
\begin{equation}
	\ket{\psi_g}=T(g)\ket{\psi_0}\quad g\in G.
\end{equation}
The states generated in this manner are not necessarily physically distinct, as some may only differ by a phase. The isotropy subgroup $H\subseteq G$ of $\ket{\psi_0}$ consists of all $h\in G$ such that 
\begin{equation}
	T(h)\ket{\psi_0}=e^{i\alpha(h)}\ket{\psi_0}.
	\label{isotropy}
\end{equation}
If $g_1,g_2\in G$ satisfy $g_2^{-1}g_1\in H$ then $\ket{\psi_{g_1}}$ and $\ket{\psi_{g_2}}$ will therefore be related by a phase and correspond to the same point on the manifold of rays $\widetilde{\ket{\psi_g}}$. To eliminate this redundancy one should consider, instead of $G$, the factor space
\begin{equation}
	X=G/H=\{gH\,:\,g\in G\}
\end{equation}
of left coset classes of $G$ with respect to $H$. Each coset class corresponds to a unique ray and so it is natural to label the coherent states using elements of $X$ rather than $G$. Let $x(g)=gH$ denote the coset class containing $g$ and let $g(x)$ denote a particular representative member of $x$. We now define the set of coherent states by
\begin{equation}
	\ket{x}=\ket{\psi_{g(x)}}\quad x\in X
\end{equation}
for which
\begin{equation}
	\ket{\psi_g}=e^{i\alpha(g)}\ket{x(g)}
\end{equation}
where $\alpha(g)$ coincides with $\alpha(h)$ on $H$. It is well known that $X$ is a homogeneous space on which $G$ acts smoothly via $g_1(gH)=(g_1 g)H$. This action is realised on the state level as
\begin{equation}
	T(g_1)\ket{x}=e^{-i\alpha(g)}\ket{\psi_{g_1 g}}=e^{i\alpha(g_1 g)-i\alpha(g)}\ket{x(g_1 g)}=e^{i\tilde{\beta}(g_1,x)}\ket{g_1 x}
	\label{csaction}
\end{equation}
where $x=x(g)$. Here $\tilde{\beta}(g_1,x)=\alpha(g_1 g)-\alpha(g)$ depends on $x$ but not the specific choice of $g$. The inner product of two coherent states will obey
\begin{equation}
	\inp{x_1}{x_2}=e^{i\tilde{\beta}(g,x_2)-i\tilde{\beta}(g,x_1)}\inp{g x_1}{g x_2}.
	\label{csinp}
\end{equation}
Comparing \eref{csaction} and \eref{csinp} with \eref{generalaction} and \eref{generalinp} reveals that the action of $G$ leaves both the metric $g$ and two-form $\sigma$ defined in \eref{provost3} invariant. In particular, the isometry group of the coherent state manifold will contain $G$ as a subgroup.

\subsection{Remarks}
\label{sectionremarks}
Before we continue it is worthwhile to remark on the dual role that will be played by the overlap $\inp{s'}{s}$ between elements of the state manifold. In the constructions that follow, $\inp{s'}{s}$ will be seen to correspond to a transition amplitude, and therefore has the nature of a two-point correlation function. The dynamical symmetries of the system at hand will be reflected as specific transformation properties of $\inp{s'}{s}$, as expressed in \eref{generalinp}. These may severely restrict, and in some cases completely determine, the functional form of $\inp{s'}{s}$. On the geometric level $\inp{s'}{s}$ acts as the ``potential'', quite literally in the K\"ahler case, from which the metric $g$ and two-form $\sigma$ are derived. The dynamical symmetries now manifest as isometries which, in turn, impose restrictions on the form of $g$. It is in this way that $\inp{s'}{s}$ will provide the link between the system's quantum mechanical dynamical symmetries and their geometric realisation on the state manifold. 

\subsection{Quantum expectation values and flows on the manifold}
\label{sectionflows}
\subsubsection{Background}
In \cite{ashtekar_geometrical_1997} Ashtekar and Schilling developed a geometric formulation of quantum mechanics in terms of the K\"ahler structure defined on the full (projective) Hilbert space. In this setting every unitary transformation $\hat{U}=\exp[i\lambda \hat{G}]$ gives rise to a transformation that leaves the metric and symplectic form  invariant. In particular, it was shown that if the expectation value of the generator $\hat{G}$ is regarded as a real scalar field on this manifold its Hamiltonian vector field is precisely the Killing vector field associated with $\hat{U}$. Here we reproduce this result with two modifications. First, since $\mathcal{S}$ is generally a submanifold of the full Hilbert  space we only consider transformations that leave $\widetilde{\scal}$ invariant, at least infinitesimally. Secondly, we allow for non-Hermitian generators $\hat{G}$, having in mind applications to conformal transformations later.

\subsubsection{Preliminaries}
Here we restrict the discussion to states which, when unnormalized, can be parametrized holomorphically by a set of $m$ complex coordinates as \mbox{$\rket{z}=\rket{z_1,\ldots,z_m}$} with $z_j=a_j+i b_j$. The tangent space at $\rket{z}$ can then be identified with ${\rm span}_{\mathbb{R}}\{\partial_{a_i}\rket{z},\partial_{b_i}\rket{z}\}$. Note that this real vector space is invariant under multiplication by $i$ in that $i\partial_{a_i}\rket{z}=\partial_{b_i}\rket{z}$. In arbitrary real coordinates $\{s_{i=1,\ldots,2m}\}$ this is expressed through a linear mapping $J$ (the complex structure \cite{moroianu_lectures_2004}) acting on the tangent space as
\begin{equation}
	i\partialsi{k}\rket{s}=J^j_{\ k}\partialsi{j}\rket{s}.
\end{equation}
In these coordinates the components of $J$ are real and satisfy $J^i_{\ k}J^k_{\ j}=-\delta^i_j$, reflecting the fact that $J^2=-1$.\\

At this stage it is useful to revisit the construction of $g$ and $\sigma$ in section \ref{sectionProvost} in a slightly different light. Consider a state $\ket{s}$ on the manifold. The tangent space at $\ket{s}$ in the \emph{full} Hilbert space $\mathcal{H}$ may be identified with $\mathcal{H}$ itself. If $\ket{a}$ and $\ket{b}$ are two tangent vectors at $\ket{s}$ then the inner product of their components orthogonal to $\ket{s}$ is
\begin{equation}
	I_{s}(\ket{a},\ket{b})=\inp{a}{b}-\inp{a}{s}\inp{s}{b}.
\end{equation}
The real and imaginary parts of $I_{s}$ are
\begin{equation}
	G_s(\ket{a},\ket{b})={\rm Re}[I_{s}(\ket{a},\ket{b})]\quad{\rm and}\quad S_s(\ket{a},\ket{b})={\rm Im}[I_{s}(\ket{a},\ket{b})]
	\label{Idef}
\end{equation}
and these are then real valued, $\mathbb{R}$-bilinear forms on this \emph{full} tangent space. These forms play a central role in the definition of the Fubini-Study metric and associated symplectic form on the projective Hilbert space \cite{chruscinski_geometric_2012}. When restricted to the tangent space of $\scal\subseteq\hcal$ these reproduce the definitions of $g$ and $\sigma$ as
\begin{equation}
	G_s(\partialsi{i}\ket{s},\partialsi{j}\ket{s})=g_{ij}\quad{\rm and}\quad S_s(\partialsi{i}\ket{s},\partialsi{j}\ket{s})=\sigma_{ij}.
	\label{GSdef}
\end{equation}
Furthermore, it is seen that $G_s(i\ket{a},\ket{b})=S_s(\ket{a},\ket{b})$ and therefore
\begin{equation}
	\sigma_{ij}=J^k_{\ i}\,g_{kj}\quad{\rm and}\quad J_{ij}=\sigma_{ji}.
	\label{sigmaidentities1}
\end{equation}
Combining the above with \eref{generalkahler} also reveals that in complex coordinates
\begin{equation}
	J_a^{\ \bar{b}}=J_{\bar{a}}^{\ b}=0\quad {\rm and}\quad J_a^{\ b}=-J_{\bar{a}}^{\ \bar{b}}=i\delta_a^b.
	\label{Jidentities1}
\end{equation}

\subsubsection{Construction}
\label{sectionflowconstruction}
Consider an operator $\hat{G}$ that generates an infinitesimal transformation akin to \eref{generalaction} that leaves the manifold of rays $\widetilde{S}$ invariant while modifying the state vectors by a scalar prefactor. There then exists a scalar function $\phi(s)$ and vector field $X_{\hat{G}}=k^i\partialsi{i}$ such that 
\begin{equation}
	\hat{G}\ket{s}=\phi(s)\ket{s}-iX_{\hat{G}}\ket{s}.
	\label{Gaction}
\end{equation}
The goal here is to relate $X_{\hat{G}}$ to $\expv{\hat{G}}\equiv\matens{s}{\hat{G}}{s}$ which can be regarded as a scalar field on the state manifold. Here $\hat{G}$ need not be Hermitian, and so we write $\hat{G}=\hat{G}_1+i\hat{G}_2$ with $\hat{G}_{1,2}$ being Hermitian. Combining \eref{Gaction} with \eref{GSdef} yields $S_s(i\hat{G}\ket{s},\ket{s}_j)=k^i\sigma_{ij}$ where $\ket{s}_j\equiv\partialsi{j}\ket{s}$. By the $\mathbb{R}$-bilinearity of $S_s$ we also have
\begin{eqnarray}
	k^i\sigma_{ij}=S_s(i\hat{G}\ket{s},\ket{s}_j)&=S_s(i\hat{G}_1\ket{s},\ket{s}_j)+S_s(-\hat{G}_2\ket{s},\ket{s}_j)\\
	&=S_s(i\hat{G}_1\ket{s},\ket{s}_j)+J^k_{\ j}S_s(-i\hat{G}_2\ket{s},\ket{s}_k)\\
	&=S_s(i\hat{G}_1\ket{s},\ket{s}_j)+J^k_{\ j}G_s(\hat{G}_2\ket{s},\ket{s}_k)\\
	&=-\frac{1}{2}\partialsi{j}\expv{\hat{G}_1}+\frac{1}{2}J^k_{\ j}\,\partialsi{k}\expv{\hat{G}_2}
\end{eqnarray}
where the final line follows from the definitions in \eref{Idef} and \eref{GSdef}. Using $\sigma_{ik}\sigma^{kj}=-\delta_i^j$ and $g^{ij}=\sigma^{ik}J_k^{\ j}$ we can express $k^i$ as
\begin{equation}
	k^i=-\frac{1}{2}\sigma^{ij}\left[\partial_j\expv{\hat{G}_1}+J_j^{\ k}\partial_k\expv{\hat{G}_2}\right]=-\frac{1}{2}\left[\sigma^{ij}\partial_j\expv{\hat{G}_1}+g^{ij}\partial_j\expv{\hat{G}_2}\right].
	\label{vectorfieldreal}
\end{equation}
The vector field $X_{\hat{G}}$ is therefore a combination of gradients associated with the symplectic and Riemannian structures respectively. Using \eref{Jidentities1} this result simplifies in complex coordinates to
\begin{equation}
	k^a=-\frac{1}{2}\sigma^{a\bar{b}}\bar{\partial}_b\expv{\hat{G}}\quad{\rm and}\quad k^{\bar{a}}=-\frac{1}{2}\sigma^{\bar{a}b}\partial_b\overline{\expv{\hat{G}}}
	\label{vectorfieldcomplex}
\end{equation}
while $X_{\hat{G}}=k^a\partial_a+k^{\bar{a}}\partial_{\bar{a}}$ then $X_{i\hat{G}}=i(k^a\partial_a-k^{\bar{a}}\partial_{\bar{a}})$.\\

\subsubsection{Vector field actions}
From \eref{Gaction} it follows that the action of $X_{\hat{G}}$ on an arbitrary expectation value $\expv{\hat{\mathcal{O}}}\equiv\matens{s}{\hat{\mathcal{O}}}{s}$ is
\begin{equation}
	X_{\hat{G}}\expv{\hat{\mathcal{O}}}=i\expv{\hat{\mathcal{O}}\delta\hat{G}-\delta\hat{G}^\dag\hat{\mathcal{O}}}
\end{equation}
with $\delta{\hat{G}}=\hat{G}-\expv{\hat{G}}$. For Hermitian $\hat{G}$ we therefore find
\begin{equation}
	X_{\hat{G}}\expv{\hat{\mathcal{O}}}=\expv{[-i\hat{G},\hat{\mathcal{O}]}}\quad{\rm and}\quad X_{\hat{iG}}\expv{\hat{\mathcal{O}}}=\expv{\{-\delta\hat{G},\hat{\mathcal{O}\}}}.
	\label{vectorfieldactionhermitian}
\end{equation}
Suppose the dynamical symmetries of $\mathcal{S}$ are generated by a Lie algebra spanned by the Hermitian operators $\{\hat{D}_i\}$. The vector fields $\{X_{\hat{D}_i}\}$ then provide a representation of the real Lie algebra spanned by $\{-i\hat{D}_i\}$ while the scalar fields $\{\expv{\hat{D}_i}\}$ transform under the adjoint representation.

\section{One- and two-dimensional state manifolds}
\label{section1d2d}
\subsection{One dimension}
\label{section1d}
As a first example we consider the one-dimensional manifold corresponding to the trajectory of a state through the Hilbert space under the reverse time evolution generated by a Hamiltonian $\hop$. We set
\begin{equation}
	\ket{t}=e^{it\hop}\ket{\phi_0}\quad{\rm with}\quad t\in\mathbb{R}.
	\label{metric1d}
\end{equation}
In the coherent state language $G=(\mathbb{R},+)$, while the isotropy subgroup $H\subseteq G$ will depend on the details of the Hamiltonian and the reference state. Excluding the case where $\ket{\phi_0}$ is a $\hop$ eigenstate, $H$  will either be trivial or, if the dynamics are periodic, a discrete subgroup of $G$. In the latter case $X=G/H\sim U(1)$ and $t$ is a periodic coordinate. The inner product from which the metric is obtained via \eref{provost3} takes the form of a transition amplitude $\inp{t}{t'}=\matens{\phi_0}{e^{-i(t-t')\hop}}{\phi_0}$. A simple calculation yields the  static metric
\begin{equation}
	g=\matens{\phi_0}{(\delta\hop)^2}{\phi_0} dt^2
	\label{1dmetric}
\end{equation}
where $\delta\hop=\hop-\matens{\phi_0}{\hop}{\phi_0}$. The distance scale in $g$ is therefore set by the uncertainty in the energy of the evolving state; a result clearly reminiscent of a time-energy uncertainty relation \cite{anandan_geometry_1990}.

\subsection{Two dimensions}
\label{section2d}
\subsubsection{Motivation}
\label{section2dmotivation}
Imagine attempting to repeat the construction of the previous section using an \emph{non-normalisable} (infinite norm) reference state $\rket{\phi_0}$. The notion of symmetry transformations which leave the resulting set $\scal_\infty\equiv\{\rket{t}\}$ invariant, as expressed in \eref{generalaction}, would still be applicable in this context. However, it would no longer be possible to construct a metric from the definitions in \eref{provost3} due to the singular nature of the inner products and matrix elements involved. Despite this difficulty, this case will be of particular interest here. This stems from the potentially enlarged set of dynamical symmetries associated with $\scal_\infty$. To see this, let us first return to the normalisable case and note that the associated metric in \eref{1dmetric} is maximally symmetric since it is one dimensional and has time translation as a continuous isometry. However, from the discussion in section \ref{generalisometries} we know that any continuous symmetry of $\scal=\{\ket{t}\}$ will also translate into an isometry, and so we must conclude that time translation can be the only continuous dynamical symmetry of $\scal$. This restriction falls away for $\scal_\infty$, since here the metric is not defined. This raises the possibility that an appropriate non-normalisable reference state could allow for a larger set of the system's dynamical symmetries to be realised as symmetries of $\scal_\infty$. Our goal will be to use these states as a starting point for constructing a manifold of \emph{normalisable} states with a corresponding metric on which this enlarged set of transformations act as isometries. It is clear that this will require both the inclusion of additional dimensions as well as a regularisation scheme for rendering the elements of $\scal_\infty$ normalisable. Crucially, the combined effect of these two operations must preserve the symmetries exhibited by $\scal_\infty$. These questions are central to our work. In the next section we show how these issues can be resolved in a simple and natural way which links closely with ideas from the $AdS_2/CFT_1$ context.\\

We remark that, given the rather restricted nature of highly symmetric two-dimensional geometries, the preceding discussion was perhaps overly general. However, the goal was to emphasise the main ideas and considerations which will also enter in higher dimensional constructions, but where the resolution of these issues are much more involved. 

\subsubsection{Construction}
\label{construction2d}
Consider again the one-dimensional construction in section \ref{section1d}. Although the following will focus on the case of a non-normalisable reference state, many of the results also apply to the normalisable case. Our task is to introduce a regularisation scheme for the states $\scal_\infty=\{\rket{t}\}$ that will allow for the definition of a metric and symplectic form while retaining the full set of dynamical symmetries exhibited by $\scal_\infty$. To accomplish this we introduce a regularisation coordinate $\beta$, thereby also increasing the dimension of the state manifold. The simplest approach for doing so also turns out to be the most useful: we complexify time and identify $\beta$ with its imaginary part. This leads to the family of states
\begin{equation}
	\rket{t,\beta}=e^{i(t+i\beta)\hop}\rket{\phi_0} \quad{\rm with}\quad (t,\beta)\in\mathbb{R}\times(\beta_0,\infty)
	\label{tbstate}
\end{equation}
where it is assumed that the $e^{-\beta\hop}$ factor renders $\rket{t,\beta}$ normalisable for $\beta>\beta_0$. Through an appropriate redefinition of the reference state we may take $\beta_0=0$. This leads to a natural partitioning of the state manifold into infinite norm ``boundary'' states at $\beta=0$ and finite norm ``bulk'' states at $\beta>0$. The metric and symplectic form can now be defined in the standard way within the bulk. The dynamical symmetries which acted on the boundary states as $\hat{U}\rket{t}\propto\rket{u(t)}$ now transform the $\rket{t,\beta}$ states according to the conformal mapping $t+i\beta\rightarrow u(t+i\beta)$ and therefore represent symmetries of both the metric and the symplectic form. These transformations also leave the $\beta=0$ boundary invariant. Furthermore, the regularisation coordinate $\beta$ clearly has the character of an energy scale and the metric therefore encodes information about the system's dynamics at different energies. These observations suggest that $\beta$ plays a role analogous to that of the radial bulk coordinate in the standard $AdS/CFT$ picture.\\

Calculating $g$ and $\sigma$ within the bulk using \eref{provost3} is straightforward, and the results can be expressed compactly in terms of the three functions
\begin{equation}
	Z(\beta)=\rinp{t,\beta}{t,\beta},\quad F(\beta)=\log Z(\beta) \quad\text{and}\quad C(\beta)=\frac{1}{4}F''(\beta)=\matens{\beta}{(\delta\hop)^2}{\beta}
	\label{zfcdef}
\end{equation}
 where $\ket{\beta}$ is $e^{-\beta H_0}\rket{\phi_0}$, properly normalised. We find
\begin{equation}
	g=C(\beta)[dt^2+d\beta^2]\quad\text{and}\quad\sigma=2C(\beta)dt\wedge d\beta
	\label{2dgands}
\end{equation}
where $C(\beta)$ is refered to as the conformal factor. The geometric information is therefore encoded in the derivatives of $F(\beta)$, which are the cumulants of $\hop$ with respect to the $\ket{\beta}$. For example, the scalar curvature is
\begin{equation}
	\mathcal{R}(\beta)=-\frac{\partial_\beta^2\log[C(\beta)]}{C(\beta)}=4\left[\gamma(\beta)^2-\kappa(\beta)\right]
\end{equation}
where $\gamma(\beta)$ and $\kappa(\beta)$ are respectively the skewness and kurtosis of the distribution of energies in $\ket{\beta}$.\\ 

Finally, we note that this is also a K\"ahler manifold with $\tau=t+i\beta$ as a natural complex coordinate. The metric and symplectic form may be calculated directly from the K\"ahler potential $\frac{1}{2}\log\rinp{\tau}{\tau}$, with $\rket{\tau}\equiv\rket{t,\beta}$, using the expressions in \eref{generalkahler}.

\subsubsection{Asymptotic behaviour}
\label{sectionaads}
The infinite norm of the reference state $\rket{\phi_0}$ implies that $Z(\beta)=\rinp{t,\beta}{t,\beta}$ will diverge as $\beta\rightarrow 0$. We are interested in the asymptotic behaviour of the metric that results from this. Taking as an ansatz $Z(\beta)=\beta^{-p}\exp[f(\beta)]$ with $p>0$ and $f(\beta)$ analytic at $\beta=0$ yields
\begin{equation}
	C(\beta)=\frac{p}{4\beta^2}+\frac{f^{(2)}(0)}{4}+\oforder{\beta}\quad\text{and}\quad\mathcal{R}(\beta)=-\frac{8}{p}-\frac{16 f^{(3)}(0)}{p^2}\beta^3+\oforder{\beta^4}.
\end{equation}
As $\beta$ approached zero the conformal factor $C(\beta)$ therefore diverges like $\sim\beta^{-2}$ while the scalar curvature approaches a constant negative value of $-8/p$. We recognise this as an (Euclidean) asymptotically Anti-de Sitter space \cite{cadoni_asymptotic_1999}. It is striking that this particular form results quite generically from this simple construction.

\subsubsection{Scale-invariance}
It is clear that the symmetries of the Hamiltonian and reference state are combined in $Z(\beta)=\rinp{t,\beta}{t,\beta}$ and imprint of the resulting metric. It is possible that these symmetries may restrict the form of $Z(\beta)$ to such an extent that the metric is completely determined. For example, suppose $\hop$ and $\rket{\phi_0}$ are invariant under scale transformations generated by a dilation operator $\hat{D}$. Specifically, 
\begin{equation}
	e^{-i\lambda \hat{D}}\hop\,e^{i\lambda \hat{D}}=e^{-\alpha\lambda}\hop\quad\text{and}\quad e^{i\lambda \hat{D}}\rket{\phi_0}=e^{\gamma\lambda}\rket{\phi_0}
\end{equation}
with $\alpha$ and $\gamma$ the two scaling dimensions. This implies that $Z(\beta)\propto\beta^{-2\gamma/\alpha}$ which by \eref{zfcdef} fixes the metric as
\begin{equation}
g=\frac{\gamma}{2\alpha\beta^2}(dt^2+d\beta^2).
\end{equation}
This is precisely $AdS_2$ with a constant scalar curvature of $\mathcal{R}=-4\alpha/\gamma$. This situation will be encountered again in section \ref{sectionCQM} in the context of conformal quantum mechanics. 
\subsubsection{Average energy as a radial coordinate}
\label{rascoordinate}
Earlier we noted the interpretation of the bulk/regularisation coordinate $\beta$ as an inverse energy scale. For later comparison with results in the literature it will be useful to employ, in the place of $\beta$, the energy expectation value $\matens{\beta}{\hop}{\beta}$ itself as a coordinate. We define
\begin{equation}
	r=-\frac{1}{2}F'(\beta)=\matens{\beta}{\hop}{\beta}
	\label{rdeff}
\end{equation}
 which is a monotonically decreasing function of $\beta$. Transforming from $(t,\beta)$ to $(t,r)$ in \eref{2dgands} yields the general forms of $g$ and $\sigma$ as
\begin{equation}
	g=\bar{C}(r)dt^2+\frac{dr^2}{4\bar{C}(r)}\quad{\rm and}\quad \sigma=-dt\wedge dr
	\label{metricitor}
\end{equation}
where $\bar{C}(r)=C(\beta)$ is the variance of $\hop$ expressed as a function of its expectation value. This is known as the \emph{variance function} in statistics literature. Now $\mathcal{R}(r)=-4\partial^2_r\bar{C}(r)$ and manifolds of constant scalar curvature are therefore associated with quadratic variance functions. To understand the dependence of $\bar{C}(r)$ on $r$ we again employ the ansatz $Z(\beta)=\beta^{-p}e^{f(\beta)}$ discussed in section 
\ref{sectionaads}. This is appropriate for an infinite norm reference state and was seen to lead to asymptotically $AdS$ geometries. Inserting $Z(\beta)$ into \eref{zfcdef} and \eref{rdeff} then yields
\begin{equation}
 C(\beta)=\frac{p}{4\beta^2}+\frac{f''(\beta)}{4}\quad{\rm and}\quad r(\beta)=\frac{p}{2\beta}-\frac{f'(\beta)}{2}
\end{equation}
where, as before, $f(\beta)$ is assumed to be analytic at $\beta=0$. At small $\beta$ and large $r(\beta)$ the variance function will therefore depend on $r$ quadratically as
\begin{equation}
	\bar{C}(r)=\frac{r^2}{p}+f'(0)\frac{r}{p}+\cdots
\end{equation}
In the opposite limit $r(\beta)$ will approach a constant value, say $r_0$, which corresponds to the energy of the lowest eigenstate of $\hop$ which has a non-zero overlap with the reference state $\ket{\phi_0}$. Since $C(\beta)\propto r'(\beta)$ this implies that $\bar{C}(r_0)=0$. The range of $r$ is therefore $(r_0,\infty)$.

\subsection{Non-static metrics}
\label{sectionsourcedmetrics}
The preceding construction can be generalised to include time-dependent Hamiltonians. We will consider the form $\hop(t)=\hop_0+\gamma(t) \hop_1$ with $\hop_0$ and $\hop_1$ static and $\gamma(t)$ a scalar function. Definition \eref{tbstate} is now modified to read
\begin{equation}
	\rket{t,\beta}=\left\{\begin{array}{rr} \hat{U}^\dag(t,0)e^{-\beta \hop_0}\rket{\phi_0} & t\geq0 \\ \hat{U}(0,t)e^{-\beta \hop_0}\rket{\phi_0} & t<0 \end{array}\right.
	\label{statedef}
\end{equation}
where $\hat{U}(t',t)$ is the time evolution operator satisfying $\partial_{t'} \hat{U}(t',t)=-i \hop(t') \hat{U}(t',t)$ and $\partial_{t} \hat{U}(t',t)=i \hat{U}(t',t)\hop(t)$. The overlap of two manifold states is a transition amplitude, now incorporating the modified dynamics due to $\gamma(t)$:
\begin{equation}
	\rinp{t',\beta'}{t,\beta}=\rmate{\phi_0}{e^{-\beta' \hop_0} \hat{U}(t',t) e^{-\beta \hop_0}}{\phi_0}\hs{1.5}(t'>t)
\end{equation}
The procedure for calculating the metric and symplectic form remains unchanged and yields
\begin{equation}
	\matcc{g_{tt} & g_{t\beta}}{g_{\beta t} & g_{\beta \beta}}=\matcc{\matens{\beta}{(\delta \hop(t))^2}{\beta} & \frac{i}{2}\matens{\beta}{\comm{\hop(t)}{\hop_0}}{\beta}}{\frac{i}{2}\matens{\beta}{\comm{\hop(t)}{\hop_0}}{\beta} & \matens{\beta}{(\delta \hop_0)^2}{\beta}}
	\label{metric2dgeneral}
\end{equation}
\begin{equation}
	\matcc{\sigma_{tt} & \sigma_{t\beta}}{\sigma_{\beta t} & \sigma_{\beta \beta}}=\matcc{0 & \frac{1}{2}\matens{\beta}{\acomm{\delta \hop(t)}{\delta \hop_0}}{\beta}}{-\frac{1}{2}\matens{\beta}{\acomm{\delta \hop(t)}{\delta \hop_0}}{\beta} & 0}
	\label{symp2dgeneral}
\end{equation}

\section{Conformal Quantum Mechanics}
\label{sectionCQM}
\subsection{Background}
In this section we apply the construction outlined in section \ref{section2d} to a model of conformal quantum mechanics introduced by de Alfaro et. al. \cite{alfaro_conformal_1976}. This model has attracted attention as a possible candidate for the one-dimensional conformal field theory dual to $AdS_2$ \cite{chamon_conformal_2011} and also appears naturally in the boundary dynamics of two-dimensional \emph{AdS} gravity \cite{cadoni_2d_2001}. It will be shown how these results can be understood within the state manifold picture.

\subsection{Definitions}
\label{sectionsu11}
The starting point for the study of conformal quantum mechanics is the classical action  \cite{alfaro_conformal_1976}
\begin{equation}
	S=\int dt \left(\dot{x}^2+\frac{g}{x^2}\right)
\end{equation}
for a single degree of freedom $x=x(t)$. Here $g\geq0$ is a coupling constant. This action is invariant under the  conformal transformations corresponding to time translations, dilations and special conformal transformations. Collectively these are realised as
\begin{equation}
	x(t)\rightarrow\tilde{x}(\tilde{t})=\sqrt{\frac{d\tilde{t}}{dt}}\,x(t)\quad{\rm with}\quad\tilde{t}=\frac{\alpha t+\beta}{\gamma t+\delta}\quad{\rm and}\quad \alpha\delta-\beta\gamma=1.
\end{equation}
On the quantum mechanical level the central objects are the three conformal generators
\begin{equation}
	\ttop=\frac{1}{2}\left(\pop^2+\frac{g}{\xop^2}\right), \quad \dop=-\frac{1}{4}\left(\xop\pop+\pop\xop\right)\quad{\rm and}\quad \kop=\frac{1}{2}\xop^2
	\label{su11generators}
\end{equation}
expressed here in terms of $\hat{x}$ and $\hat{p}$ which satisfy $[\hat{x},\hat{p}]=i$. These form a representation of the $su(1,1)\cong so(2,1)$ algebra:
\begin{equation}
	[\ttop,\dop]=i\ttop\quad\quad\quad[\kop,\dop]=-i\kop\quad\quad\quad[\ttop,\kop]=2i\dop.
\end{equation}
The $su(1,1)$ Casimir operator $\hat{C}=\frac{1}{2}[\ttop\kop+\kop\ttop]-\dop^2$ is found to equal $k(k-1)\hat{I}$ with $k=(1+\sqrt{g+1/4})/2$ and the coupling constant $g$ therefore determines the particular $su(1,1)$ irrep under consideration. The Hilbert space is spanned by the states $\{\ket{k,n}\ :\ n=0,1,2,\ldots\}$ which satisfy
\begin{align}
\kz\ket{k,n}&=(k+n)\ket{k,n}\nonumber\\
\kp\ket{k,n}&=\sqrt{(n+1)(2k+n)}\ket{k,n+1}\label{su11basis}\\
\km\ket{k,n}&=\sqrt{n(2k+n-1)}\ket{k,n-1}\nonumber
\end{align}
where
\begin{equation}
	\kz=\frac{1}{2}[\kop+\ttop]\quad{\rm and}\quad \kop_\pm=\frac{1}{2}[\kop-\ttop]\pm i\dop.
	\label{su11basis2}
\end{equation}
Appendix \ref{appendixa} collects various algebraic identities for this algebra.

\subsection{Dynamical symmetries and geometry}
Here we will consider the dynamics generated by $\ttop$ and set $\ket{t}=e^{it\ttop}\ket{\phi_0}$. Other choices for the Hamiltonian will be discussed in section \ref{otherhamiltonians}. The question is now how the reference state $\ket{\phi_0}$ should be chosen to ensure that $\{\ket{t}\}$ exhibits the full set of $su(1,1)$  symmetries. The discussion of coherent states in section \ref{sectionGCS} suggests seeking a reference state which is invariant under dilations and special conformal transformations, i.e. a simultaneous eigenstate of $\kop$ and $\dop$. However, no such state exists in the Hilbert space of normalisable states, and we are forced to widen our search to include infinite norm states as well. The only candidate for $\rket{\phi_0}$ is 
\begin{equation}
	\rket{\phi_0}=e^{-\kop_+}\ket{k,0}
	\label{cqmreferencestate}
\end{equation}
which has infinite norm and satisfies $\dop\rket{\phi_0}=-ik\rket{\phi_0}$ and $\kop\rket{\phi_0}=0$. This can be verified using the expressions in \eref{su11basis}, \eref{su11basis2} and \eref{su11identity1}. The corresponding family of states $\{\rket{t}=e^{it\ttop}\rket{\phi_0}\}$ is precisely that studied in \cite{chamon_conformal_2011}. The dynamical symmetries are realised as
\begin{equation}
	e^{i\lambda \hop}\rket{t}=\rket{t+\lambda}\quad\quad e^{i\lambda \dop}\rket{t}=e^{\lambda k}\rket{e^\lambda t} \quad\quad e^{i\lambda \kop}\rket{t}=(1-\lambda t)^{-2k}\rkets{\textstyle \frac{t}{1-t\lambda}}
	\label{tstatetransform}
\end{equation}
which, as per \eref{generalinp}, imply certain transformation properties for the inner product $\rinp{t'}{t}$ and restricts its form accordingly. It is found that
\begin{equation}
	\rinp{t'}{t}=\left[\frac{i}{2(t-t')}\right]^{2k}
	\label{tinnerproduct}
\end{equation}
which has the form of a two-point functions of a field with conformal dimension $k$ \cite{chamon_conformal_2011}. The expressions in \eref{tstatetransform} and \eref{tinnerproduct} above follow from combining \eref{su11basis} and \eref{su11basis2} with \eref{su11identity2} to \eref{su11identity4}. Due to the divergent behaviour of $\rinp{t'}{t}$ at $t'=t$ no geometry can be defined on this set of states. We therefore proceed as in section \ref{construction2d} and replace $t\rightarrow\tau=t+i\beta$ and study $\rket{\tau}\equiv\rket{t,\beta}=e^{i\tau \ttop}\rket{\phi_0}$. The transformation equations in \eref{tstatetransform} remain valid after this replacement, while the inner product becomes
\begin{equation}
	\rinp{\tau'}{\tau}=\left[\frac{i}{2(\tau-\bar{\tau}')}\right]^{2k}.
	\label{tauinnerproduct}
\end{equation}
In particular, $\rinp{\tau}{\tau}=(4\beta)^{-2k}$ is finite for $\beta>0$. The resulting metric is
\begin{equation}
	ds^2=\frac{k}{2\beta}\left[dt^2+d\beta^2\right]
\end{equation}
which is precisely $AdS_2$ with a scalar curvature of $\mathcal{R}=-4/k$. The three isometries of this metric correspond to the three dynamical symmetries expressed in \eref{tstatetransform}. Through the regularisation the symmetries of the infinite norm boundary states have therefore been translated into isometries of the bulk geometry. Note that these transformations leave the $\beta=0$ boundary invariant. The nature of the bulk states become clear upon noting that, by \eref{su11identity2},
\begin{equation}
	\rket{\tau}=e^{i\tau \ttop}\ket{\phi_0}\propto e^{z(\tau)\kp}\ket{k,0}
\end{equation}
where the state on the right is, up to normalisation, just the standard $SU(1,1)$ coherent state with parameter $z(\tau)=(\tau-i)/(\tau+i)$ \cite{perelomov_generalized_1986}. This reveals that the full set of states \mbox{$\{\rket{\tau=t+i\beta}\ :\ (t,\beta)\in\mathbb{R}\times[0,\infty)\}$} therefore consists of two disjoint families of coherent states: the infinite norm boundary states with $\beta=0$ and $|z(\tau)|=1$ and the normalisable bulk states with $\beta>0$ and $|z(\tau)|<1$. 

\subsection{General $su(1,1)$ Hamiltonians}
\label{otherhamiltonians}
In the previous section we selected $\hat{H}=\hat{T}$ as the generator of time evolution, but in principle any other $su(1,1)$ algebra element with a spectrum bounded from below could also serve as Hamiltonian. However, this will not lead to a distinct set of states or a different geometry. For example, the Hamiltonian $\hat{H}_{ho}=\ttop+\alpha^2 \kop$ is the analogue of a harmonic oscillator and possesses an equidistant spectrum. According to \eref{su11identity2} and \eref{su11identity3} the states $\rket{\tau}_{ho}=e^{i\tau(\ttop+\alpha^2 \kop)}\rket{\phi_0}$ are related to those of the ``free particle'' Hamiltonian $\ttop$ through the conformal mapping $\tau_{fp}=\tan(\alpha\tau_{ho})/\alpha$. In particular, the underlying state manifold and geometric structures are identical in these two cases. It is perhaps surprising that the geometry generated using $\hat{H}_{ho}$, which is clearly not scale invariant, exhibits the same isometries as that associated with $\ttop$. This underscores that the geometry is really determined by the dynamical symmetry group, rather than the particular element of the algebra chosen as the Hamiltonian. The same observations apply to non-static metrics generated from time-dependent Hamiltonians via the procedure outlined in section \ref{sectionsourcedmetrics}. If we restrict ourselves to $su(1,1)$ generators the resulting bulk geometry must be identical to the $su(1,1)$ coherent state geometry, although parametrised in a more complicated way. 

\subsection{The generators of conformal transformations on the operator and geometric levels}
Here we apply the notions developed in section \ref{sectionflows} on the link between quantum expectation values and flows on the manifold to the $su(1,1)$ model of conformal quantum mechanics. For concreteness we develop the following in terms of the $\rket{\tau}=e^{i\tau\ttop}\rket{\phi_0}$ states. However, as outlined in section \ref{otherhamiltonians}, the main results will be applicable to any $su(1,1)$ Hamiltonian.  In section \ref{sectionflowconstruction} a link was established between the expectation value of an operator $\hat{G}$ and the vector field $X_{\hat{G}}$ of the infinitesimal transformation it generates on the state manifold. Here we use this result to investigate the generators of conformal transformations, of which the dynamical symmetry generators constitute a subset, on the operator and geometric levels.\\

It has been seen that the dynamical symmetries of $\scal=\{\rket{\tau}\}$ are generated by the $su(1,1)$ algebra spanned by $\{\ttop,\kop,\dop\}$. These act on the unnormalised $\rket{\tau}$ states as
\begin{equation}
	\ttop\rket{\tau}=-i\partial_\tau\rket{\tau},\quad \dop\rket{\tau}=-i(k+\tau\partial_\tau)\rket{\tau}\quad{\rm and}\quad\kop\rket{\tau}=-i(2k\tau+\tau^2\partial_\tau\rket{\tau},
	\label{su11diffop}
\end{equation}
which are just the infinitesimal and complexified versions of \eref{tstatetransform}. This suggests a natural extension to the full set of conformal generators defined by
\begin{equation}
	\vop_n\rket{\tau}=-i\left[(n+1)k\tau^n+\tau^{n+1}\partial_\tau\right]\rket{\tau}
	\label{vndef}
\end{equation}
where $\vop_1=\kop$, $\vop_{-1}=\ttop$ and $\vop_0=\dop$. It can be verified that $\{i\vop_n\}$ satisfies the centreless Virasoro algebra:
\begin{equation}
	[i\vop_n,i\vop_m]=(n-m)i\vop_{n+m}.
\end{equation}
In fact, these operators can be realised explicitly in terms of $su(1,1)$ subalgebra as
\begin{eqnarray}
	i\vop_n&=(i\vop_0+nk)\frac{\Gamma(i\vop_0+k)}{\Gamma(i\vop_0+k+n)}(i\vop_1)^n\\
	i\vop_{-n}&=(i\vop_0-nk)\frac{\Gamma(i\vop_0+1-k-n)}{\Gamma(i\vop_0+1-k)}(i\vop_{-1})^n
\end{eqnarray}
where $n\geq0$ \cite{fairlie_construction_1988}. The expectation value of $\vop_n$ can be calculated by combining \eref{vndef} with the explicit form of $\rinp{\tau}{\tau}$ in \eref{tauinnerproduct}. Inserting this expectation value into \eref{vectorfieldcomplex} then yields the conformal vector fields
\begin{equation}
	X_{\vop_n}=\tau^n\partial_\tau+\bar{\tau}^n\partial_{\bar{\tau}}\quad{\rm and}\quad X_{i\vop_n}=i(\tau^n\partial_\tau-\bar{\tau}^n\partial_{\bar{\tau}}).
\end{equation}
The Killing vector fields $\{X_{\vop_{\pm1}},X_{\vop_{0}}\}$  provide a representation of $su(1,1)$ acting on the space of scalar fields defined on $\scal$. For fields that are the expectation values of operators the transformations properties under this representation will obviously mimic those of the operators themselves. For example, consider the conformal generators defined above. For $|n|>1$ these generators are not Hermitian, and so we isolate their Hermitian and anti-Hermitian parts as $\vop_m=\vop_m^{(1)}+i \vop_m^{(2)}$. Using \eref{vectorfieldactionhermitian} the corresponding set of real expectation values are found to satisfy
\begin{equation}
	X_{\vop_n}\expv{\vop_m^{(1,2)}}=\expv{[-i\vop_n,\vop_m^{(1,2)}]}=-(n-m)\expv{\vop_{m+n}^{(1,2)}}
\end{equation}
for $n\in\{0,\pm1\}$ and $m\in\mathbb{Z}$ and therefore carries a representation of $su(1,1)$. The Casimir operator \mbox{$\hat{C}=X_{\vop_0}^2-(X_{\vop_1}X_{\vop_{-1}}+X_{\vop_{-1}}X_{\vop_{1}})/2$} then satisfies
\begin{equation}
	\hat{C}\expv{\vop_m^{(1,2)}}=2\expv{\vop_m^{(1,2)}}\quad\quad m\in\mathbb{Z}.
\end{equation}
Despite the constancy of $\hat{C}$ this representation is not irreducible and may be decomposed as follows:
\begin{itemize}
	\item The subspaces spanned by $\{\expv{\vop^{(2)}_{n}}\}_{n>1}$ and $\{\expv{\vop^{(2)}_{n}}\}_{n<1}$ are invariant and carry the $k=2$ positive and negative discrete series representations of $su(1,1)$ respectively.
	\item The subspace spanned by $\{\expv{\vop_{0,\pm1}}\}$ is invariant and carries the adjoint representation of $su(1,1)$. This may also be regarded as the $j=1$ representation of $su(2)$ via the association $\hat{J}_z\sim X_{\vop_0}$, $\hat{J}_+\sim X_{\vop_1}$ and $\hat{J}_-\sim -X_{\vop_{-1}}$. 
	\item The subspaces spanned by $\{\expv{\vop^{(1)}_{n}}\}_{n>1}$ and $\{\expv{\vop^{(1)}_{n}}\}_{n<1}$ are not invariant. When restricted to these subspaces the representation is equivalent to the $k=2$ positive and negative discrete series representations of $su(1,1)$ respectively. Although reducible, this representation is therefore not completely reducible.
\end{itemize}
The expectation values of elements of the enveloping algebra of $su(1,1)$ can be classified in a similar way. In fact, through the association with $su(2)$ outlined above this amounts to the construction of spherical tensor operators. For example, starting with $\expv{\vop_1\vop_1}=\expv{\kop^2}$ and applying $X_{\vop_{-1}}$ repeatedly generates a set of five real fields which transform under the $j=2$ representation of $su(2)$ and are eigenfunctions of $\hat{C}$ with eigenvalue $2(2+1)$. 

\subsubsection{Summary}
Here we collect the results above in a form appropriate for use in later sections. It was shown that the real scalar fields on $\scal$ may be classified according to their transformation properties with respect to the $su(1,1)$ Killing vector fields. This leads to the eigenvalue equation $\hat{C}\phi(\tau,\bar{\tau})=\lambda\phi(\tau,\bar{\tau})$. As is well known $\hat{C}$ is closely related to the Laplace-Beltrami operator. Specifically,
\begin{equation}
	\nabla^2=-\frac{\mathcal{R}}{2}\hat{C}=-\frac{\mathcal{R}}{2}(\tau-\bar{\tau})^2\partial_\tau\partial_{\bar{\tau}}
\end{equation}
with $\mathcal{R}=-4/k$ and the eigenvalue equation can therefore be expressed as
\begin{equation}
	\nabla^2\phi(\tau,\bar{\tau})=-\frac{\mathcal{R}}{2}\lambda\phi(\tau,\bar{\tau}).
	\label{laplaceeq}
\end{equation}
For $\lambda=2$ the solutions for $\phi(\tau,\bar{\tau})$ are linear combinations of the real and imaginary parts of the expectation values of the conformal generators.\\

The vector field associated with a generator $\hat{G}$ is $X_{\hat{G}}=k^\tau\partial_\tau+k^{\bar{\tau}}\partial_{\bar{\tau}}$ where, according to \eref{vectorfieldcomplex}, $k^\tau=-\frac{1}{2}\sigma^{\tau\bar{\tau}}\partial_{\bar{\tau}}\expv{\hat{G}}$. For this to be a conformal transformation $k^\tau$ must be holomorphic. Indeed, the requirement that $\partial_{\bar{\tau}}k^\tau=0$ is equivalent to the conformal Killing equation 
\begin{equation}
	\nabla_a k_b+\nabla_b k_a=(\nabla\cdot k)g_{ab}.
	\label{conformalkilling}
\end{equation}
 If we regard this as an equation for $\expv{\hat{G}}$ instead of for $\{k^\tau,k^{\bar{\tau}}\}$ then the general complex solution is a linear combination of $\{\expv{\vop_n}\ :\ n\in\mathbb{Z}\}$ plus an arbitrary holomorphic function. However, the only \emph{real} solutions are linear combinations of $\expv{\vop_{0,\pm1}}$ plus an arbitrary constant.\\

Although we have focused on the states $\rket{\tau}=e^{i\tau\ttop}\rket{\phi_0}$ the results regarding the solutions to equations \eref{laplaceeq} and \eref{conformalkilling} are applicable to any $su(1,1)$ Hamiltonian. 

\section{State manifolds as solutions of classical dilaton gravity}
\label{sectiondilaton}
\subsection{Introduction}
The geometries we have considered thus far have all emerged from a quantum mechanical setting in which inner product, reference state and representation of the dynamical symmetry group all featured explicitly. We now turn to the other side of the duality and ask whether these geometries, and the scalar fields defined on them, can be seen as the result of a classical gravitational theory. In particular, this will involve interpreting equations \eref{laplaceeq} and \eref{conformalkilling}, which are essentially algebraic in nature, as the equations of motion of a scalar field.
\subsection{Static metrics}
\subsubsection{General}
In section \ref{section2d} we considered two-dimensional geometries which result from time complexification. The form of the metric for this case appeared in \eref{2dgands} and \eref{metricitor}. The key observation here is that this form of $g$ appears quite generally as the solution of classical dilaton gravity in two dimensions. The action of a range of such dilaton models can be brought into the form \cite{cadoni_trace_1996,cadoni_asymptotic_1999}
\begin{equation}
	S[g,\eta]=\frac{1}{2\pi}\int\ dx\sqrt{g}[\eta \mathcal{R}+V(\eta)]
	\label{dilatonaction}
\end{equation}
where $\eta$ is a scalar field related to the dilaton, $V(\eta)$ an arbitrary potential and $\mathcal{R}\equiv\mathcal{R}[g]$ the scalar curvature. For conciseness we will refer to $\eta$ itself as the dilaton. Varying with respect to $g$ and $\eta$ yields the equations of motion
\begin{equation}
	\mathcal{R}[g]=-V'(\eta)\quad{\rm and}\quad\nabla_\mu\nabla_\nu\eta-\frac{1}{2}g_{\mu\nu}V(\eta)=0.
\end{equation}
The general static solution for $g$ and $\eta$ takes the form
\begin{equation}
	g=\bar{C}(r)dt^2+\frac{dr^2}{4\bar{C}(r)}\quad{\rm and}\quad \eta(t,r)=r
\end{equation}
where $g$ is exactly as in \eref{metricitor} and $\bar{C}(r)$ is related to the potential by $V(r)=4\bar{C}'(r)$. In other words, the dilaton action that produces a metric with a particular variance function $\bar{C}(r)$ has a potential which depends on $\eta$ as $4\bar{C}'(r)$ depends on $r$. Being static this metric exhibits time translation as an isometry, which is generated by the Hamiltonian on the operator level and by the Killing field $\partial_t$ on the geometric level. The link here is the results of \eref{vectorfieldreal} and \eref{metricitor} which together imply that $X_{\hat{H}}=\partial_t$, as it should be. As will be seen in more detail in the next section, the fact that the solution for $\eta$ coincides with the expectation value of a generator of an isometry is no coincidence. In the dilaton context the interest usually falls on metrics which are asymptotically $AdS$, which requires that $\bar{C}(r)\sim r^2$ at large $r$. As was seen in section \ref{rascoordinate} this asymptotic behaviour is exactly what is found in state manifolds based on infinite norm reference states.

\subsubsection{$AdS_2$}
Let us consider now a maximally symmetric example of the preceding construction, one which is based on the model of conformal quantum mechanics introduced in section \ref{sectionCQM}. We take as Hamiltonian $\hop=\hat{T}+\alpha^2\hat{K}$ and consider the state manifold of $\rket{\tau}=e^{i\tau\hop}\rket{\phi_0}$ with $\rket{\phi_0}$ as in \eref{cqmreferencestate}. For finite $\alpha$ the dynamics is periodic and $t\in[0,\frac{\pi}{\alpha})$. The lower bound on $r=\expv{\hop}$ is found to be $r_0=2\alpha k$ which is the groundstate energy of $\hop$. In $(t,r)$ coordinates the metric reads
\begin{equation}
	g=\frac{r^2-r_0^2}{2k}dt^2+\frac{2k}{r^2-r_0^2}\frac{dr^2}{4}
\end{equation}
which has a constant scalar curvature of $\mathcal{R}=-4/k$. As discussed in section \ref{otherhamiltonians} this metric is again just $AdS_2$ and therefore maximally symmetric. According to the preceding discussion this metric solves the equations of motion for the dilaton model in \eref{dilatonaction} for a linear potential $V(\eta)=\frac{4}{k}\eta$. This is the Jackiw-Teitelboim model \cite{Teitelboim,jackiw_lower_1985}, perhaps the most studied model of two dimensional gravity. As before $\eta(t,r)=r$ is a solution for the scalar field equation of motion. However, now other solutions also exist, a fact closely tied to the enhanced symmetry of this geometry. To see this, we first note that the equation of motion for $\eta$ now reads 
\begin{equation}
	\nabla_\mu\nabla_\nu\eta+\frac{1}{2}g_{\mu\nu}\mathcal{R}\eta=0.
	\label{dilatoneom}
\end{equation}
For the two dimensional case at hand it can be verified that these equations are completely equivalent to 
\begin{equation}
	\nabla^2\eta+\mathcal{R}\eta=0\quad{\rm and}\quad\nabla_\mu k_\nu+\nabla_\nu k_\mu=0
	\label{dilatoneom2}
\end{equation}
where $k^\mu=-(1/2)\sigma^{\mu\nu}\nabla_\nu\eta$. In particular, the latter condition states that $k^\mu$ must be a Killing vector field. According to the discussion following equation \eref{conformalkilling}, $\eta$ must therefore be, up to a constant, a linear combination of the expectation values of $\ttop$, $\kop$ and $\dop$. The unknown constant is fixed at zero by the first equation in \eref{dilatoneom2} which requires that $\eta$ is a linear combination of the real and imaginary parts of the expectation values of the conformal generators $\{\vop_n\}$. This follows from the discussion after equation \eref{laplaceeq}. We conclude that there are three linearly independent solutions for $\eta$, namely $\expv{\ttop}$, $\expv{\kop}$ and $\expv{\dop}$. 

In the gravitational context it is known that there is associated with a solution of \eref{dilatoneom} the \emph{constant} quantity
\begin{equation}
	M=-\frac{1}{2}\left[(\nabla\eta)^2+\mathcal{R}\eta^2/2\right]
	\label{mdef}
\end{equation}
which is traditionally identified with the mass or energy of the gravitational system \cite{mann_conservation_1993,navarro_generalized_1997,navarro_symmetries_1997}. One expects the value of $M$ to reflect an intrinsic property of the generator $\hat{G}=u\ttop+v\dop+w\kop$ corresponding to the particular solution $\eta=\expv{\hat{G}}$. Evaluating \eref{mdef} yields $M=\frac{k}{2}(4uw-v^2)$. The algebraic content of this quantity can be understood by noting that $(2v^2-8uw)$ is the norm squared of $iG$ (as an abstract algebra element) with respect to the $su(1,1)$ Killing form. Positive $M$ therefore corresponds to compact operators, negative $M$ to non-compact hyperbolic operators and vanishing $M$ to parabolic operators \cite{alfaro_conformal_1976}. 

\subsection{Quantum and dilaton dynamics}
In this final section we investigate the link between the dynamics of the symmetry generators on the quantum level and the dynamics of the dilaton itself. For concreteness we again focus on a particular choice of $su(1,1)$ Hamiltonian, namely
\begin{equation}
	\hop(t)=\ttop+\gamma(t)\kop=\frac{1}{2}\left(\pop^2+\frac{g}{\xop^2}\right)+\frac{\gamma(t)}{2}\xop^2.
\end{equation}
Here $\gamma(t)$ is an arbitrary time-dependent source. Returning now to $(t,\beta)$ coordinates we find using \eref{metric2dgeneral} and \eref{symp2dgeneral} that
\begin{equation}
	ds^2=\frac{k}{2\beta^2}\left[(1-\gamma(t)\beta^2)^2 dt^2+d\beta^2\right]\quad{\rm and}\quad\sigma=\frac{k(1-\gamma(t)\beta^2)}{\beta^2}dt\wedge d\beta.
\end{equation}
The metric is still $AdS_2$ with a scalar curvature of $\mathcal{R}=-4/k$. As before the solution to the dilaton equations of motion is a linear combination of the expectation values of $\{\ttop,\dop,\kop\}=\{\vop_{-1},\vop_0,\vop_1\}$, now with respect to the states in \eref{statedef}. On the quantum mechanical level the dynamics of the Heisenberg picture generators are given by
\begin{equation}
	\frac{d}{dt}\vop_n(t)=[i\hop_H(t),\vop_n(t)]=\sum_{k=-1}^1\Lambda_{kn}\vop_k(t) \quad{\rm with}\quad \Lambda=\matccc{0 & -1 & 0}{2\gamma(t) & 0 & -2}{0 & \gamma(t) & 0}.
	\label{Veom}
\end{equation}
Here $\Lambda$ is $iH_H(t)$ in the adjoint representation of $su(1,1)$. The general solution for the dilaton reads $\eta(t,\beta)=\matens{t,\beta}{\hat{G}}{t,\beta}=\matens{\beta}{U(t)\hat{G}U^\dag(t)}{\beta}$ where $U(t)\hat{G}U^\dag(t)=\sum_i \alpha_i(t)\vop_i$ and $\hat{G}\in su(1,1)$. It is straightforward to show that the $\alpha_i$-coefficients satisfy
\begin{equation}
	\frac{d}{dt}\alpha_n(t)=-\sum_{k=-1}^1\Lambda_{nk}\vop_k(t).
	\label{Aeom}
\end{equation}
Note that \eref{Veom} and \eref{Aeom} differ through the replacement of $\Lambda$ with $-\Lambda^T$. Whereas the evolution of the generators is governed by $iH_H(t)$ in the adjoint representation, the $\alpha_i$-coefficients evolve according to $iH_H(t)$ in the \emph{dual} of the adjoint representation. However, for semi-simple Lie algebras these two representations are known to be isomorphic, in this case through the mapping $\alpha_{\pm 1}\Leftrightarrow \vop_{\mp 1}$ and $\alpha_{0}\Leftrightarrow -2\vop_{0}$. Indeed, this mapping transforms the equations of motion for the $\alpha_i$-coefficients in \eref{Aeom} into those for the generators in \eref{Veom}. Returning to dilaton solution $\eta(t,\beta)$ we find
\begin{equation}
	\eta(t,\beta)=\matens{t,\beta}{\hat{G}}{t,\beta}=\matens{\beta}{\sum_i \alpha_i(t)\vop_i}{\beta}=\frac{k\alpha_{-1}(t)}{\beta}+k\alpha_1(t)\beta
	\label{etaform}
\end{equation}
in which $\alpha_0(t)$ does not appear explicitly. Here we have made use of \eqref{tauinnerproduct} and \eqref{su11diffop} to show that $\matens{\beta}{\hop}{\beta}=k/\beta$, $\matens{\beta}{\kop}{\beta}=k\beta$ and $\matens{\beta}{\dop}{\beta}=0$. Eliminating $\alpha_0(t)$ and $\vop_0(t)$ from \eref{Veom} and \eref{Aeom} then yields the two sets of equations of motion
\begin{eqnarray}
	\ddt{\alpha_1}&=-\gamma(t)\ddt{\alpha_{-1}}\hs{2}   &\ddtt{\alpha_{-1}}=-2\gamma(t)\alpha_{-1}+2\alpha_{1}\label{Aeom2}\\
	\ddt{\ttop}&=-\gamma(t)\ddt{\kop}\hs{2}   &\ddtt{\kop}=-2\gamma(t)\kop+2\ttop
\end{eqnarray}
It is now clear that the dynamics of $\alpha_1$ and $\alpha_{-1}$ exactly mirror that of $\ttop=\frac{1}{2}\left(\pop^2+\frac{g}{\xop^2}\right)$ and $\kop=\frac{1}{2}\xop^2$. Indeed, it can be verified that \eref{Aeom2} is exactly the result of substituting the form of $\eta(t,\beta)$ on the right of \eref{etaform} into the dilaton equations of motion in \eref{dilatoneom}. These observations agree with those of \cite{cadoni_2d_2001} which identify the boundary dynamics of 2D $AdS$ gravity with that of a conformally invariant mechanical system. Here we see how this comes about in a concrete way. The quantum dynamics are encoded into the state manifold via the definitions in \eref{statedef} and are then transferred to the metric in \eref{metric2dgeneral}. This metric is a solution of the 2D dilaton gravity model, and this dynamic information is then encoded in the dynamics of the dilaton and, through the self-duality of the adjoint representation, appear in the equations of motion for the expansion coefficients of $\eta(t,\beta)$.
\section{Discussion}
We have analysed various aspects of the procedure outlined in section \ref{sectionintroduction} and demonstrated its implementation for the case of two dimensional state manifolds generated through complexified time evolution. The central theme was that the geometric structures defined on the quantum Hilbert space can be used to generate, in a systematic and constructive manner, manifolds with geometries which encode a desired set of quantum mechanical symmetries. Furthermore, the geometries that result from this, together with a certain class of scalar fields defined on them, allowed for an interpretation from the viewpoint of two-dimensional dilaton gravity. Several aspects of the $AdS/CFT$ philosophy, such as the emergence of additional bulk dimensions and the generic asymptotically $AdS$ nature of the metric, followed from this approach. In this way we were able to provide a novel perspective on existing results in the $AdS_1/CFT_2$ literature. Also noteworthy was the important role played by the symplectic structure; a quantity which is usually absent in standard implementations of the $AdS/CFT$ duality.\\

Many questions and avenues for further investigation are evident. Higher dimensional constructions for systems exhibiting conformal or Schr\"odinger symmetry can be performed, and preliminary results \cite{van_zyl_constructing_2015}  show this this approach indeed generates  the geometries introduced in previous studies of gravity duals for non-relativistic $CFT$s \cite{son_toward_2008,mcgreevy_holographic_2010}. However, in this higher dimensional setting translating between the quantum mechanical and gravitational interpretations of the geometry, and the structures defined on it, will be more challenging. We will pursue these questions elsewhere. Also of interest would be a better understanding of how the generating functional for correlation functions, the central quantity in higher dimensional $AdS/CFT$ constructions, can be incorporated in the two-dimensional case. The path integral formalism for coherent state transition amplitudes might provide insight here \cite{kochetov_path_1995}. Although we have not focused on this aspect here, a detailed study of the physical content of the state manifold's Riemannian curvature would also be interesting. Results in this direction have appeared in \cite{provost_riemannian_1980}.
\label{sectionconclusion}

\acknowledgments
HJRvZ gratefully acknowledges the financial support of the Wilhelm Frank Trust.

\appendix
\section{BCH and related identities for $su(1,1)$}
\label{appendixa}
In the discussion of the $su(1,1)$ model of $CQM$ we make use of several identities of the Baker-Campbell-Hausdorff variety. These are easily generated using symbolic computation and the two dimensional irrep of $su(1,1)$ \cite{gilmore2012lie}:
\begin{equation}
	\hop=\frac{1}{2}\matcc{1 & -1}{1 & -1}\quad\quad\dop=-\frac{i}{2}\matcc{0 & 1}{1 & 0}\quad\quad\kop=\frac{1}{2}\matcc{1 & 1}{-1 & -1}.
\end{equation}
For example, it can be verified that
\begin{equation}
e^{\hat{K}_+}(u\ttop+v\dop+w\kop)e^{-\hat{K}_+}=(2u-iv)\hat{K}_0-2u\hat{K}_++\frac{1}{2}[w+iv-u]\hat{K}_-.
\label{su11identity1}
\end{equation}
Products of exponentials involving algebra elements can also be brought into the canonical form $\exp[Y_+\hat{K}_+]\exp[2\log[Y_0]\hat{K}_0]\exp[Y_-\hat{K}_-]$. The following cases will prove to be useful:
\begin{align}
	e^{i\tau\ttop}e^{-\hat{K}_+}&\quad\Longrightarrow\quad Y_+=\frac{\tau-i}{\tau+i} &&Y_0=\frac{i}{i+\tau}\label{su11identity2}\\
e^{i\lambda\dop}e^{i\tau\ttop}e^{-\hat{K}_+}&\quad\Longrightarrow\quad Y_+=1+\frac{2}{i \tau e^\lambda-1} &&Y_0=\frac{ie^{\lambda/2}}{i+\tau e^\lambda}\\
e^{i\lambda\kop}e^{i\tau\ttop}e^{-\hat{K}_+}&\quad\Longrightarrow\quad Y_+=-1+\frac{2i\tau}{\tau(i+\lambda)-1} &&Y_0=\frac{1}{1-\tau(i+\lambda)}\\
	e^{i\tau(\ttop+\alpha^2\kop)}e^{-\hat{K}_+}&\quad\Longrightarrow\quad Y_+=\frac{\tan(\alpha\tau)/\alpha-i}{\tan(\alpha\tau)/\alpha+i} \quad\quad&& Y_0=\frac{\alpha}{\alpha\cos(\alpha\tau)-i\sin(\alpha\tau)}\label{su11identity3}
\end{align}	
Finally, for $\exp[\bar{z}\hat{K}_-]\exp[z\hat{K}_+]$ we find $Y_0=(1-z\bar{z})^{-1}$ and so
\begin{equation}
		\matens{k,0}{e^{\bar{z}\hat{K}_-}e^{z\hat{K}_+}}{k,0}=(1-z\bar{z})^{-2k}.
		\label{su11identity4}
\end{equation}

\bibliographystyle{jhep}
\bibliography{adscqm}
\end{document}